# Effect of Temperature and TiO$_2$ NP Concentration on DNA Stability under Conditions Close to Physiological Ones

Evgeniya Usenko[1], Anastasiia Svidzerska[1], Alexander Glamazda[1,2*], Vladimir Valeev[1], Anna Laguta[2,3], and Victor Karachevtsev[1]

[1]B. Verkin Institute for Low Temperature Physics and Engineering of the National Academy of Sciences of Ukraine, 47 Nauky Ave., Kharkiv 61103, Ukraine

[2]V. N. Karazin Kharkiv National University, 4 Svobody sq., Kharkiv 61022, Ukraine

[3]Aston University, School of Infrastructure and Sustainable Engineering, Department of Chemical Engineering and Applied Chemistry, Energy and Bioproducts Research Institute, Birmingham B4 7ET, UK

\* Corresponding author:

Dr. Alexander Glamazda,

B. Verkin Institute for Low Temperature Physics and Engineering of the National Academy of Sciences of Ukraine, 47 Nauky Ave., Kharkiv 61103, Ukraine

e-mail: glamazda@ilt.kharkov.ua

Evgeniya Usenko (0000-0002-8446-5504 – ORCID ID)

Anastasiia Svidzerska (0000-0001-5675-5611 – ORCID ID)

Alexander Glamazda (0000-0003-3048-8732 – ORCID ID)

Vladimir Valeev (0009-0008-6659-7280 – ORCID ID)

Anna Laguta (0000-0002-0736-2923 – ORCID ID)

Victor Karachevtsev (0000-0003-4580-6465 – ORCID ID)

**Abstract**

The present work is devoted to the study of the thermostability of native DNA during binding to $TiO_2$ nanoparticles (NPs) at various concentrations under conditions close to physiological (0.1 M $Na^+$, pH 7) using thermal denaturation method and dynamic light scattering (DLS). The analysis of the DNA melting curves in the presence of $TiO_2$ NPs revealed a temperature range in which the light absorption of DNA decreases. This is explained by the formation of an ordered structure of partially unwound DNA strands on the NP surface. The DNA:$TiO_2$ NP nanoassemblies formed at the initial stages remained stable in the wide temperature range. The performed temperature-dependent DLS measurements of the DNA:$TiO_2$ NP suspension at pH 7 have not revealed the aggregation of the DNA:$TiO_2$ NP nanoassemblies that had been observed at pH 5.

**Keywords**

$TiO_2$ nanoparticles, DNA, nanoassemblies, thermal denaturation, dynamic light scattering

## Introduction

The stability of the DNA molecular structure plays a main role in the functioning of a living organism. Viruses, drugs, and environmental factors can inhibit DNA replication. At the moment, it is already known about the ambiguous effect of nanoparticles on the structural stability of DNA and in particular $TiO_2$ nanoparticles (NPs), which are widely used in various spheres of human life [1,2]. For example, $TiO_2$ NPs are quite intensively used in environmental engineering, cosmetology, pharmaceuticals, and medicine [3-8]. But, despite the growing demand for these NPs in production and medicine, numerous works are currently appearing, indicating their possible negative impact on human health, as well as their genotoxicity [9-10]. For example, in Refs. [11-14] the effect of $TiO_2$ NPs on DNA was studied by spectroscopic methods. The data presented in the work [12] indicate both the presence of electrostatic interaction of positively charged $TiO_2$ NPs with negatively charged phosphate groups of DNA and the formation of chemical bonds between $TiO_2$ NPs and DNA. This interaction leads to disruption of the secondary structure of the biopolymer. In Ref. [15] the molecular mechanisms of DNA damage in HepG2 cells induced by $TiO_2$ NPs were studied. It was shown that $TiO_2$ NPs can affect gene expression in DNA.

In the work [11] fluorescence spectra were obtained for DNA complexes with $TiO_2$ NPs. The binding constant of $TiO_2$ NPs to DNA was ~$4.2 \times 10^6$ $M^{-1}$ [11], which indicates that NPs have a high affinity for DNA and can directly bind to the biopolymer. In addition, it has been reported that NPs can strongly inhibit DNA replication and change the conformation of polynucleotides, which can lead to genotoxicity [11].

The formation of DNA:$TiO_2$ NP complexes was studied in Refs. [13,14,16-21]. It has been established that the adsorption of DNA on NPs is affected by the experimental conditions (temperature, pH), and biopolymer length [16], the NPs' concentration [13]. In particular, studies [13] have shown that both the formation of complexes between DNA and individual NPs and the formation of large conglomerates [14] consisting of several NPs bound by a polymer are possible. In addition, Refs. [13,14] show that temperature intensifies the formation of such complexes.

Despite the already available experimental data, some questions remain open. In particular, it is of interest to find out how the concentration of NPs affects a DNA conformation under conditions close to physiological ones. Also of interest is the temperature effect on the interaction of DNA with $TiO_2$ NPs. In particular, in most of the studies performed in neutral solutions, short fragments of DNA or RNA and a fixed temperature were mainly used for research [16-21]. In the recent Refs. [13,14], devoted to the study of the effect of temperature on the interaction between DNA and TiO$_2$ NPs, the measurements were performed at pH 5. The thermal studies are an analogue of the action of DNA helicase promoting the unwinding of the duplex DNA during the replication. Therefore, the study of the effect of temperature on the stability of DNA in the presence of various agents, including nanoparticles, is of significant interest.

In the present work, for the first time, the effect of TiO$_2$ NPs on the thermal stability of native DNA under conditions close to physiological (0.1 M Na$^+$, pH 7) was studied. This concentration of sodium ions is close to the total concentration of monovalent ions in the intracellular fluid [22], therefore, the study of complex formation between TiO$_2$ NPs and DNA under such ionic conditions is of greatest interest. Moreover, biofunctionalized nanoparticles can serve as model objects for studying their biocompatibility with various biological objects and appropriate drug delivery strategies.

**Materials and Methods**

The TiO$_2$ NP powder purchased from Sigma-Aldrich (particle diameter (d) < 100 nm (BET), Product code: 677469) was used in the present work. TiO$_2$ NPs were dispersed in distilled water with a pH of 6.65 which is close to the isoelectric point [23]. The NP suspension was obtained by ultrasonication with $\nu$ = 22 kHz for 40 min at room temperature. The complexes of TiO$_2$ NPs with DNA were prepared as follows: (i) the salmon sperm DNA (M$_w$ = (4 – 6) × 10$^6$ Da) purchased from Serva (Germany) was added to a buffer solution with 10$^{-3}$ M sodium cacodylate (CH$_3$)$_2$AsO$_2$Na•3H$_2$O from Serva (Germany), 0.099 M NaCl at pH 7; (ii) the required number of the TiO$_2$ NPs was added to the DNA buffer suspension. The DNA phosphorous concentration [P] was (7.14 ± 1) × 10$^{-5}$ M had been determined by the molar extinction coefficient at $\nu_m$ = 38500 см$^{-1}$ [24]. In the present work, a DLS study of the DNA:TiO$_2$ NP nanoassemblies was performed in the cacodylate buffer suspension (0.1 M Na$^+$, pH 7) at a molar concentration of TiO$_2$ ([c$_{TiO2}$]) of 1.5 × 10$^{-4}$ M. The concentration of polynucleotide phosphates [P] equaled 8 × 10$^{-5}$ M.

DNA thermal melting is an valuable method for the study of structural biopolymer conformations. The analysis of the thermal measurements allows us to determine some thermodynamic binding parameters of ligands to DNA, such as the melting temperature, the temperatures of the beginning and end of the helix-coil transition, and the transition interval. It is worth noting that the temperature of the DNA denaturation has found its application in medicine, for example, in a polymerase chain reaction (PCR) requiring a melting of the DNA double helix into single strands of DNA [25].

In the present work, thermal denaturation was used to study the structural stability of DNA in the presence of $TiO_2$ NPs. Upon increasing the temperature, the ordered structure of DNA becomes disturbed. As a result, the biopolymer forms loops that appear because of breaking some H-bonds between the nitrogen-paired bases. When the temperature reaches a value close to the melting temperature ($T_m$), the number of helical regions becomes equal to the number of unwound regions. In other words, $T_m$ is the temperature at which 50 % of a DNA sequence is in the helix conformation, and the other 50 % is present as single strands. A further increase in the temperature leads to a strong shift of equilibrium towards the increase of the fraction of the single strands. This process is accompanied by an increase in the UV absorption intensity. The dynamics of intensity growth decreases when the double helix of the biopolymer is completely unwound. The high molecular DNA undergoes melting in the temperature range from 3 to 20 °C. The melting curve describes the dependence of the UV absorption intensity on temperature. The melting curves of DNA with different $TiO_2$ concentrations were obtained using a UV-spectrophotometer at a fixed wavenumber of the band with $\nu_m = 38500$ см$^{-1}$ that corresponds to the maximum absorption of DNA. This band is quite wide, and its possible shift as a result of heating or an increase in the concentration of $TiO_2$ NPs lies within the experimental error of the spectral measurements. We have used the laboratory software to perform the registration of the melting curves as the temperature dependence of the hyperchromicity coefficient: $h(T)=[\Delta A(T)/A_{T_o}]_{\nu m}$, where $\Delta A(T)$ is a change in the optical density of the DNA suspension upon heating and $A_{T_o}$ is the optical density at $T = T_o$ Thus, $h(T)$ is the quantitative characteristic of hyperchromism. The registration of the absorption intensity was carried out using the double-cuvette scheme: identical suspension of DNA or their complexes with $TiO_2$ were placed in both channels of the spectrophotometer. The reference cuvette was thermostated at $T = T_0 \pm 0.5$ °C, while the sample cuvette was slowly heated at a rate of 0.25 °C/min from 20 to 94 °C.

The particle size distributions were determined via dynamic light scattering (DLS) using Zetasizer Nano ZS (Red badge) ZEN 3600 Malvern Instruments apparatus at different temperatures in the range of 25 – 90 °C (temperature range when thermal denaturation of DNA occurs). DLS measurements were carried out at 632.8 nm (output power of 4 mW) with a He-Ne laser at a 175° scattering angle. The temperature control in a cuvette holder proceeded automatically with a step 2 °C. The duration of the isotherm was 30 s (except for the measurements at 25 °C, which is the first point, the remaining measurements were carried out with an equilibrium time of 120 s). This time interval didn't include a short time for temperature regulation by the apparatus. Three particle size measurements for each temperature point were proceeded (and ten in the case of 25 °C). Every measurement consisted of at least 10 runs

(automatic choice). Every measurement is based on an average value of correlation curves, every one of which is the result of a single run [26]. Such an averaging correlation curve was treated by two approaches [27]. The first one is a cumulants' analysis which is the fit of a polynomial to the log of the correlation function and the second order cumulant includes the Z-average diffusion coefficient. The second mechanism of the treatment of correlation curves is that the correlation function is a superposition of exponential decay for each particle size in the sample. Then exponential decay parameters can be transformed into particle size distribution [27]. To assign the viscosity value was used "solvent builder", which calculated the viscosity of the dispersant depending on NaCl concentration. A glass cuvette was used in the case of size measurements (since there were high-temperature measurements).

**Results and discussion**

**Thermal denaturation**

The melting curve of DNA exhibits the usual S-like curve upon heating caused by only one spiral-coil transition (Fig. 1). The DNA melting process takes place in a fairly narrow temperature range ($\Delta T$) of about 10 °C. It indicates a high degree of cooperativity of the helix-coil structural transition. Upon the $TiO_2$ NP injection into the DNA suspension, there are no noticeable changes in the shape of the melting curve up to $[c_{TiO2}] = 2.5 \times 10^{-5}$ M. However, there is a trough appears in the shape of the melting curve upon injection of $[c_{TiO2}] = 5 \times 10^{-5}$ M. The depth (absorption hypochromism) of the trough increases with $[c_{TiO2}]$. With adding $[c_{TiO2}] = 2 \times 10^{-4}$ M, the value of h is -0.17. The several linear segments can be distinguished on each of the presented melting curves at $[c_{TiO2}] \geq 5 \times 10^{-5}$ M. They are connected to the beginning ($T_{s1}$, $T_{s2}$) and finish ($T_{f1}$, $T_{f2}$) of the definite ongoing processes discussed below. Let's examine the DNA melting curve in the presence of $[c_{TiO2}] = 2 \times 10^{-4}$ M (curve 7, Fig.1). The extrapolation of the linear segments is shown in the black dashed lines in Fig.1. A similar trough-like form of melting curves was observed earlier in acidic solutions [13]. According to Ref.[13], the absorption hypochromism observed in the segment of a-b with increasing temperature is caused by the formation of a more ordered structure of the biopolymer bound to $TiO_2$ NPs. The argument in favor of this assumption is the fact that the cooperativity of the change in h upon heating from $T_{S1}$ to $T_{f1}$ is of the same order as the cooperativity of the DNA helix–coil transition at $T > T_{s2}$ (Fig. 1). It should be noted that, in our case, the formation of DNA:$TiO_2$ NP nanoassemblies takes place already at room temperature (this fact is confirmed by the DLS method (see below)), and with increasing temperature, these nanoassemblies are formed more intensively. The unwound DNA regions bind to NP and form a more ordered structure on the NP surface than double-stranded DNA at the initial moment in the suspension with TiO$_2$ NPs. This explains the decrease in the h value. In the segment of b-c, the formed DNA:TiO$_2$ NP nanoassemblies remain stable. This is somewhat different from the studies performed at pH 5 [14], where the aggregation of DNA:TiO$_2$ NP nanoassemblies was observed over the segment of b-c. The observed hyperchromism on the melting curve at T > 70 °C (the segment of c-d) is due to the helix-coil transition.

It should be noted that the recently performed study of DNA:TiO$_2$ NP suspensions at pH 5 revealed the formation of the light scattering nanoaggregates whose size was comparable to the wavelength of visible light at a concentration of [$c_{TiO2}$] = 1.75 × 10$^{-4}$ M [14]. This was also reflected in the distortion of the S-like shape of the melting curve of double-stranded DNA. In the present study, the formation of the DNA:TiO$_2$ NP nanoaggregates was not detected up to a concentration of [$c_{TiO2}$] = 2 × 10$^{-4}$ M.

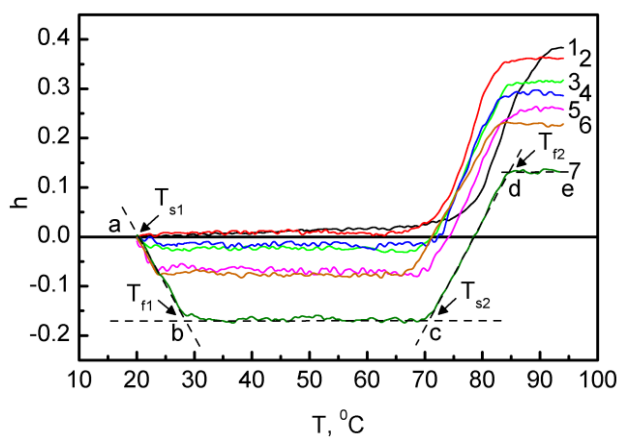

**Fig. 1** Temperature dependence of hyperchromic coefficient (h) of DNA without (curve 1) and with (curves 2-7) TiO$_2$ NPs (0.1 M Na$^+$, pH 7): «1» – [$c_{TiO2}$] = 0; «2» – [$c_{TiO2}$] = 2.5 × 10$^{-5}$ M; «3» – [$c_{TiO2}$] = 5 × 10$^{-5}$ M; «4» – [$c_{TiO2}$] = 10$^{-4}$ M; «5» – [$c_{TiO2}$] = 1.5 × 10$^{-4}$ M, «6» – [$c_{TiO2}$] = 1.75 × 10$^{-4}$ M; «7» – [$c_{TiO2}$] = 2 × 10$^{-4}$ M.

The performed analysis of the melting curves presented in Fig. 1 allowed us to obtain the dependence of the melting temperature on [$c_{TiO2}$] (Fig. 2). Figure 2 shows the $T_m$([$c_{TiO2}$]) dependences determined in the segment of c-d in Fig.1. From this figure, it is seen that the addition of [$c_{TiO2}$] = 2.5 × 10$^{-5}$ M to the DNA solution leads to a noticeable decrease in the DNA melting temperature by ~ 6 °C. The further increase in [$c_{TiO2}$] stimulates slight changes in $T_m$ (changes are within 1.9 °C). The reason for this behavior of the concentration dependence of $T_m$ will be discussed below.

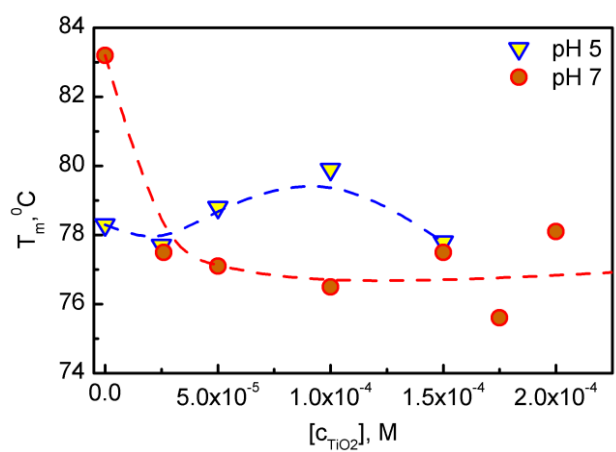

**Fig. 2** The DNA melting temperature dependence as a function of [$c_{TiO2}$] at pH 5 [13] and pH 7. The dashed lines are for the eye guideline.

As noted earlier [12], DNA can interact with $TiO_2$ NPs in the following ways: 1) electrostatic binding between $TiO_2$ NPs and the phosphate group of DNA; 2) interaction of $TiO_2$ NPs with nitrogenous bases of DNA; 3) formation of P–O–Ti covalent bonds between the phosphate backbone and the NP surface. At present, it is already quite well known that the interaction of $TiO_2$ NPs with the DNA phosphate groups increases the thermal stability of this polynucleotide, while the interaction with nitrogenous bases, on the contrary, lowers the thermal stability of DNA [13]. Thus, the reason of such behaviour of the concentration dependence of the DNA melting temperature is determined by the total contribution of all possible types of interaction between $TiO_2$ NPs and DNA. These interactions can have a different dominant character in the entire $TiO_2$ NP concentration and temperature ranges. The decrease in $T_m$ observed at [$c_{TiO2}$] $\leq 2.5 \times 10^{-5}$ M is because the main contribution to the concentration dependence of the melting temperature is caused by the interaction of $TiO_2$ NPs with nitrogenous bases of DNA. The $T_m$ dependence reflects a very weak evolution at [$c_{TiO2}$] $> 2.5 \times 10^{-5}$ M that is apparently due to compensation for the effects causing an increase and decrease in the thermal stability of DNA due to the implementation of all possible types of binding of DNA to $TiO_2$ NPs. It should be noted that in our previous studies performed at pH 5, slight changes in the melting temperature were observed in the entire studied range of [$c_{TiO2}$] [13].

It is also known that $TiO_2$ NPs become positively or negatively charged upon protonation or deprotonation of the nanoparticle surface in a solution with different pH: at pH < 6.5, NPs are positively charged, and at pH > 6.5, they are negatively charged [23]. At a pH of about 6.5, NPs have a neutral charge [23]. It follows from the foregoing that $TiO_2$ NPs studied under conditions close to physiological ones are negatively charged. Nevertheless, the high ionic strength of the solution partially compensates for the negative charge on the oxygen atoms of the phosphate groups and, accordingly, reduces the repulsion between the negatively charged TiO$_2$ NPs and the negatively charged DNA phosphate groups.

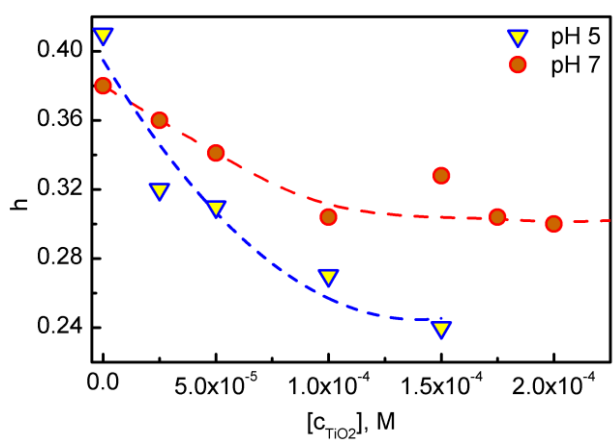

**Fig. 3** The dependence of the DNA hyperchromic coefficient (h) in the presence of TiO$_2$ NPs on [$c_{TiO2}$] at pH 5 [13] and pH 7. The dashed lines are for the eye guideline.

At the same time, at pH 7, the interaction of the nitrogenous bases of DNA with TiO$_2$ NPs is rather strong. This interaction is dominant in these ionic conditions. Unlike pH 7, TiO$_2$ NPs are positively charged at pH 5. Under these conditions, the interaction of NPs with phosphate groups and nitrogenous bases is quite pronounced. Moreover, the contribution of these interactions to the T$_m$ dependence on [$c_{TiO2}$] is approximately the same. This circumstance is the reason for the weak dependence of T$_m$ on [$c_{TiO2}$] at pH 5 [13].

The h([$c_{TiO2}$]) dependences of DNA (the values of h obtained by analysis of the melting curves in the segment of c-d in Fig.1) at pH 7 and pH 5 are plotted in Fig. 3. It can be seen from this figure that the addition of [$c_{TiO2}$] = $1.0 \times 10^{-4}$ M to the DNA suspension leads to a decrease in the value of h([$c_{TiO2}$]) by ~ 21 % at pH 7 and by ~ 34 % at pH 5. Further addition of TiO$_2$ NPs to the DNA solution practically does not change the value of this parameter. This result seems quite natural. As noted above, the decrease in the melting temperature is associated with the interaction of nitrogenous bases of DNA (in particular, the N7 atoms of DNA guanine [13]) with a surface of TiO$_2$ NPs. It leads to the appearance of unwound regions in the structure and, accordingly, to a decrease in the degree of helicity of the polynucleotide. Thus, the observed decrease in h is precisely caused by this effect. In addition, as noted above, at pH 7, TiO$_2$ NPs are negatively charged. Despite the fact that the high ionic strength of the suspension partially compensates for the negative charge on the NP surface, nevertheless, the remaining electrostatic repulsion between TiO$_2$ NPs and DNA phosphate groups weakens the interaction between NPs and DNA. Such an effect is reflected in a more gradual change in the value of h at pH 7 compared with pH 5.

**Dynamic light scattering**

In our previous studies of DNA:TiO$_2$ NP aqueous suspensions studied at pH 5, we observed the occurrence of aggregation processes near [$c_{TiO2}$] = 1.5 × 10$^{-4}$ M. The performed analysis of the DNA melting studies presented above have not revealed the aggregation in the concentration range of [$c_{TiO2}$] = 2.5 × 10$^{-5}$ – 2 × 10$^{-4}$ M at pH 7. This fact has stimulated us to perform the DLS studies of DNA:TiO$_2$ NP aqueous suspensions at [$c_{TiO2}$] = 1.5 × 10$^{-4}$ M for assessing the colloidal stability of the nanoassemblies. So, as is known, we can obtain by the three different types of size distributions: intensity- , volume-, and number-weighted distribution (from measuring the fluctuation in scattering intensity), as well as hydrodynamic diameter (Z-ave or $d_H$), which is calculated from diffusion coefficient (cumulant analysis). As we reported earlier [13], it is the most reasonable to use the DLS size distribution by number in the case of nanoparticle characterization [28-31]. However, we shouldn't avoid distributions by intensity from the DLS method because this one is initial for DLS at all. The calculation of distributions by volume and number is based on the assumption, that particles are closer to spherical shape. So, some errors can appear in these distributions. In any case, the number-based average diameter of the used TiO$_2$ NPs is 81 ± 5 nm. The distribution by intensity and by volume is slightly different, despite a relatively low polydispersity index (PdI) of 0.17 ± 0.04 and assuming the spherical shape of NPs, Z-ave = 119 nm.

Figure 4 shows the comparative analysis of cross-sections extracted from 3D size distributions by number for DNA:TiO$_2$ NP solution taken at pH 5 and pH 7 at T = 25 $^0$C. There are some minor peaks at the small diameter value in the size distribution measured at pH 7 but the main peak looks similar to that one for pure TiO$_2$ NP (pH 6.65) and DNA:TiO$_2$ NP (pH 5) colloidal solutions. The nature of the minor peaks has an artificial character not corresponding to actual existing particles [32]. They are not important for the discussion of changes in main peaks. The number-based average diameter of the DNA:TiO$_2$ NP complexes is 98 ± 13 nm. The PdI value is also slightly larger than in the case of pure TiO$_2$ NPs and is equal to 0.27 ± 0.01. Increasing in intensity-base size may be pointed at the appearance of a higher amount of DNA:TiO$_2$ NP nanoassemblies with a strong scattering of light.

As shown above, the DNA melting temperature in the absence of TiO$_2$ NPs under the studied ionic conditions is 83.5 °C. So, we can suppose that the appearance of unwounded DNA strands upon heating will stimulate an increase in the size of the DNA:TiO$_2$ NP nanoassemblies as was shown in our previous publication [14]. So, when heating to 90 °C the size distributions by number were obtained (See Fig. 5).

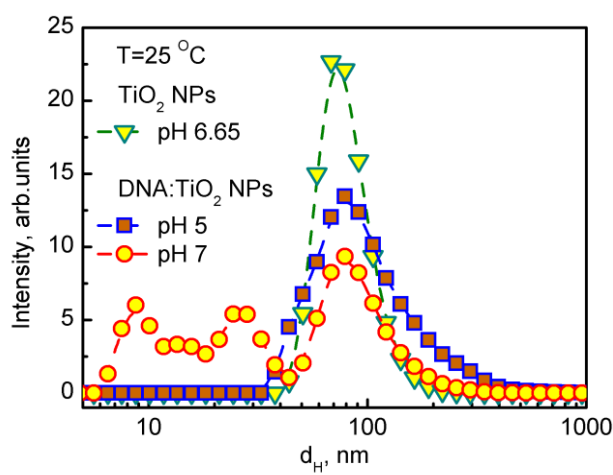

**Fig. 4** The size distributions by number for DNA:TiO$_2$ NP colloidal suspensions taken at pH 5 [14] and pH 7 (cacodylate buffer), T = 25 $^0$C.

In this case, minor peaks disappeared, however, the main peaks of distribution practically didn't change in comparison with DNA:TiO$_2$ NP solution at 25 °C.

The number-based average diameter of the DNA:TiO$_2$ NP nanoassemblies at pH 7 is 112 ± 5 nm which differs from the estimated diameter of the DNA:TiO$_2$ NP nanoaggregates measured at pH 5 equals 213 ± 8 nm. It indicates that no the DNA:TiO$_2$ NP nanoaggregates are formed at pH 7 in the entire temperature region in contrast to pH 5.

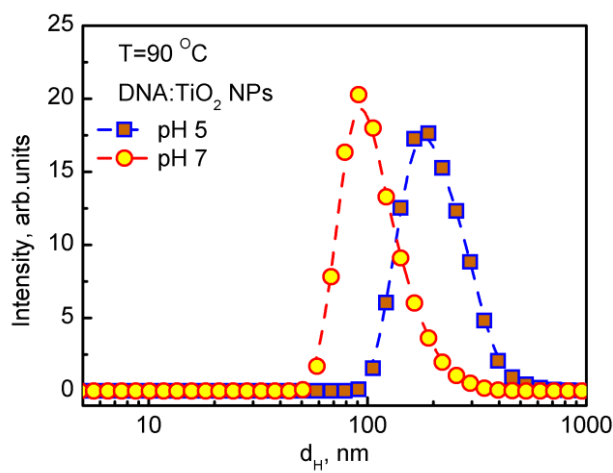

**Fig. 5** The comparative analysis of the cross-section extracted from 3D size distributions by number for DNA:TiO$_2$ NP solution taken at pH 5 and pH 7 at T = 82 – 90 °C (average values).

The PdI value of DNA:TiO$_2$ NP suspension (pH 7) is also slightly higher than in the case of pure TiO$_2$ NPs and is equal to 0.18 ± 0.02. The recently performed study of DNA:TiO$_2$ NP suspensions at pH 5 has PdI equal to 0.34 ± 0.05 (25 °C) and 0.24 ± 0.01 (90 °C). It demonstrates a slight decrease in PdI of DNA:TiO$_2$ NP suspension at pH 7.

In conclusion, we can say, that the DNA thermal melting data obtained by DLS showed that DNA:TiO$_2$ NP nanoassemblies formed at room temperature in an aqueous suspension at pH 7 remain stable in the entire studied temperature range of 25 – 90 °C. It is explained by the strong binding of DNA to the TiO$_2$ NP surface that prevents the formation of the nanoaggregates of DNA:TiO$_2$ NP nanoassemblies observed recently at pH 5 [14]. Moreover, heating intensifies the formation of nanoassemblies due to the appearance of the unwound single-stranded DNA regions at pH 7 as it was found at pH 5 [13]. In addition, TiO$_2$ NPs have a negative charge at pH 7 (at pH 5, they are positively charged), which, in turn, affects the nature of the interaction between DNA and TiO$_2$ NPs and, accordingly, is reflected in the concentration dependence of the melting temperature and hyperchromic coefficient. The results obtained in the present work are predominantly fundamental, since, on the one hand, they provide insight into the nature and mechanisms of the interaction of DNA with TiO$_2$ NPs at different pH values, and, on the other hand, they can be useful for further production of the multifunctional antibacterial coatings.

**Conclusions**

The study of the effect of temperature and TiO$_2$ NP concentration on the stability and conformation of DNA under conditions close to physiological (0.1 M Na$^+$, pH 7) was performed by UV spectroscopy, thermal denaturation, and dynamic light scattering. The analysis of the DNA melting curves in the presence of TiO$_2$ NPs revealed a temperature range in which the light absorption of DNA decreases. We believe that the observed effect is explained by the unwound DNA regions that can bind to NPs and form a more ordered structure on the NP surface than double-stranded DNA at the initial moment. The injection of [c$_{TiO2}$] = 2.5 × 10$^{-5}$ M into DNA suspension leads to a decrease in the DNA melting temperature by ~ 6°C at pH 7. It is assumed that this effect is due to the predominant interaction of the nitrogenous bases of DNA with these NPs. The performed temperature-dependent DLS measurements of the DNA:TiO$_2$ NP suspension at pH 7 have not revealed the aggregation of the DNA:TiO$_2$ NP nanoassemblies that had been observed at pH 5. From the results of our work, it is clear that TiO$_2$ NPs can significantly affect the functioning of DNA in a living organism. At the same time, we also found that the DNA:TiO$_2$ NP nanoassemblies are biocompatible and do not form aggregates under conditions close to physiological. Thus they can serve as model objects for the development of drug-related products based on TiO$_2$ NPs, for example, in photodynamic therapy, where separation and size control of acting nanoassemblies is necessary. The results obtained in this study can be used to create various hygiene products, as well as medicine and pharmacology.

**Availability of data and materials**

The datasets used and/or analysed during the current study are available from the corresponding author on reasonable request.

**Funding**

Authors acknowledge financial support from National Academy of Sciences of Ukraine (Grant No. 0120U100157).

**Ethics approval and consent to publish, participate**

We confirm that this manuscript had not been published elsewhere before and would not be considered to be published on other journals.

**Conflicts of Interest**

The authors do not have any commercial or associative interest that represents any conflict of interest in connection with the work submitted.

**Author contribution statement**

All authors discussed the results and commented on the manuscript. V. Valeev and E. Usenko carried out the spectroscopic measurements, thermal denaturation and analyzed data. A. Svidzerska and A. Laguta performed and analyzed the DLS data. A. Glamazda and V. Karachevtsev planned and coordinated the project. E. Usenko, A. Svidzerska, A. Glamazda, and V. Karachevtsev wrote the paper.

**References**

1. Baranowska-Wójcik E, Szwajgier D, Oleszczuk P, Winiarska-Mieczan A (2020) Effects of titanium dioxide nanoparticles exposure on human health-a review. Biol Trace Elem Res 193:118–129. https//doi.org/10.1007/s12011-019-01706-6
2. Shabbir S, Kulyar MF-E-A, Bhutta ZA, Boruah P, Asif, M (2021) Toxicological consequences of titanium Dioxide nanoparticles ($TiO_2$ NPs) and their jeopardy to human population. Bionanoscience 11:621–632. https//doi.org/10.1007/s12668-021-00836-3
3. Arun J, Nachiappan S, Rangarajan G, Alagappan RP, Gopinath KP, Lichtfouse E (2023) Synthesis and application of titanium dioxide photocatalysis for energy, decontamination and viral disinfection: a review. Environ Chem Lett 21:339–362. https//doi.org/10.1007/s10311-022-01503-z
4. Givelet L, Truffier-Boutry D, Noël L, Damlencourt JF, Jitaru P, Guérin T (2021) Optimisation and application of an analytical approach for the characterisation of $TiO_2$ nanoparticles in food additives and pharmaceuticals by single particle inductively coupled plasma-mass spectrometry. Talanta 224:121873. https//doi.org/10.1016/j.talanta.2020.121873
5. Włodarczyk R, Kwarciak-Kozłowska A (2021) Nanoparticles from the cosmetics and medical industries in legal and environmental aspects. Sustainability 13(11):5805. https//doi.org/10.3390/su13115805
6. Jafari S, Mahyad B, Hashemzadeh H, Janfaza S, Gholikhani T, Tayebi L (2020) Biomedical applications of $TiO_2$ nanostructures recent advances. Int J Nanomedicine 15:3447–3470. https//doi.org/10.2147/IJN.S249441
7. Musial J, Krakowiak R, Mlynarczyk DT, Goslinski T, Stanisz BJ (2020) Titanium dioxide nanoparticles in food and personal care products-What do we know about their safety? Nanomaterials (Basel, Switzerland) 10(6):1110. https//doi.org/10.3390/nano10061110
8. Ziental D, Czarczynska-Goslinska B, Mlynarczyk DT, Glowacka-Sobotta A, Stanisz B, Goslinski T, Sobotta L (2020) Titanium dioxide nanoparticles: prospects and applications in medicine. Nanomaterials (Basel, Switzerland) 10(2):387. https//doi.org/10.3390/nano10020387
9. Chang X, Zhang Y, Tang M, Wang B (2013) Health effects of exposure to nano-$TiO_2$:a meta-analysis of experimental studies. Nanoscale Res Let 8:51. https://doi.org/10.1186/1556-276X-8-51

10. Shah SN, Shah Z, Hussain M, Khan M (2017) Hazardous effects of titanium dioxide nanoparticles in ecosystem. Bioinorg Chem Appl 2017:4101735. https//doi.org/10.1155/2017/4101735

11. Patel S, Patel P, Bakshi SR (2017) Titanium dioxide nanoparticles an in vitro study of DNA binding, chromosome aberration assay, and comet assay. Cytotechnology 69:245–263. https//doi.org/10.1007/s10616-016-0054-3

12. Patel S, Patel P, Undre SB, Pandya SR, Singh M, Bakshi S (2016) DNA binding and dispersion activities of titanium dioxide nanoparticles with UV/vis spectrophotometry, fluorescence spectroscopy and physicochemical analysis at physiological temperature. J Mol Liq 213:304–311. https//doi.org/10.1016/j.molliq.201511002

13. Usenko E, Glamazda A, Valeev V, Svidzerska A, Laguta A, Petrushenko S, Karachevtsev V (2022) Effect of $TiO_2$ nanoparticles on the thermal stability of native DNA under UV irradiation. Appl Phys A 128:900. https//doi.org/10.1007/s00339-022-06043-5

14. Usenko E, Glamazda A, Svidzerska A, Valeev V, Svidzerska A, Laguta A, Petrushenko S, Karachevtsev V (2023) DNA:$TiO_2$ nanoparticle nanoassemblies: effect of temperature and nanoparticle concentration on aggregation. J Nanopart Res 25:113. https//doi.org/10.1007/s11051-023-05770-x

15. El-Said KS, Ali EM, Kanehira K, Taniguchi A (2014) Molecular mechanism of DNA damage induced by titanium dioxide nanoparticles in toll-like receptor 3 or 4 expressing human hepatocarcinoma cell lines. J Nanobiotechnology 12:48. https//doi.org/10.1186/s12951-014-0048-2

16. Zhang X, Wang F, Liu B, Kelly EY, Servos MR, Liu J (2014) Adsorption of DNA oligonucleotides by titanium dioxide nanoparticles. Langmuir 30(3):839–845. https//doi.org/10.1021/la404633p

17. Bae S, Oh I, Yoo J, Kim JS (2021) Effect of DNA flexibility on complex formation of a cationic nanoparticle with double-stranded DNA. ACS Omega 6(29):18728–18736. https//doi.org/10.1021/acsomega.1c01709

18. Cleaves HJ 2nd, Jonsson CM, Jonsson CL, Sverjensky DA, Hazen RM (2010) Adsorption of nucleic acid components on rutile ($TiO_2$) surfaces. Astrobiology 10(3):311–323. https//doi.org/10.1089/ast.2009.0397

19. Ma L, Liu B, Huang P-JJ, Zhang X, Liu J (2016) DNA Adsorption by ZnO nanoparticles near its solubility limit implications for DNA fluorescence quenching and DNAzyme activity assays. Langmuir 32(22):5672–5680. https//doi.org/10.1021/acs.langmuir.6b00906

20. He Q, Wu Q, Feng X, Liao Z, Peng W, Liu Y, Peng D, Liu Z, Mo M (2020) Interfacing DNA with nanoparticles: surface science and its applications in biosensing. Int J Biol Macromol 151:757–780. https//doi.org/10.1016/j.ijbiomac.2020.02.217
21. Das S, Chatterjee S, Pramanik S, Devi PS, Kumar GS (2018) A new insight into the interaction of ZnO with calf thymus DNA through surface defects. J Photochem Photobiol B 178:339–347. https//doi.org/10.1016/j.jphotobiol.2017.10.039
22. Saenger W (1998) Principles of nucleic acid structure. Springer, New York
23. li HM, Babar H, Shah TR, Sajid MU, Qasim MA, Javed S (2018) Preparation techniques of $TiO_2$ nanofluids and challenges: a review. Appl Sci 8(4):587. https//doi.org/10.3390/app8040587
24. Sorokin VA, Valeev VA, Usenko EL, Andrushchenko VV (2012) DNA conformational equilibrium in the presence of $Zn^{2+}$ ions in neutral and alkaline solutions. Int J Biol Macromol 50(3):854–860. https//doi.org/10.1016/j.ijbiomac.2011.11.011
25. Mullis K, Faloona F, Scharf S, Saiki R, Horn G, Erlich H (1986) Specific enzymatic amplification of DNA in vitro the polymerase chain reaction. Cold Spring Harb Symp Quant Biol 51:263–273. https//doi.org/10.1101/SQB.1986.051.01.032
26. Svanberg C, Bergman R (2020) Photon correlation spectroscopy. Definitions. https://doi.org/10.32388/hfvlwa
27. Tscharnuter W (2006) Photon correlation spectroscopy in particle sizing. In: Meyers RA and Flippen RB (eds) Encyclopedia of Analytical Chemistry. John Wiley & Sons Ltd. https://doi.org/10.1002/9780470027318.a1512
28. Cascio C, Geiss O, Franchini F, Ojea-Jimenez I, Rossi F, Gilliland D, Calzolai L (2015) Detection, quantification and derivation of number size distribution of silver nanoparticles in antimicrobial consumer products. J Anal At Spectrom 30:1255–1265. https//doi.org/10.1039/C4JA00410H
29. De la Calle I, Menta M, Klein M, Maxit B, Séby F (2018) Towards routine analysis of $TiO_2$ (nano-)particle size in consumer products: evaluation of potential techniques. Spectrochim Acta Part B At Spectrosc 147:28–42. https//doi.org/10.1016/j.sab.2018.05.012
30. Krause B, Meyer T, Sieg H, Kästner C, Reichardt P, Tentschert J, Jungnickel H, Estrela-Lopis I, Burel A, Chevance S, Gauffre F, Jalili P, Meijer J, Böhmert L, Braeuning A, Thünemann AF, Emmerling F, Fessard V, Laux P, Lampen A, Luch A (2018) Characterization of aluminum, aluminum oxide and titanium dioxide nanomaterials using a combination of methods for particle surface and size analysis. RSC Adv 8:14377–14388 https//doi.org/10.1039/C8RA00205C

31. Souza TG, Ciminelli VS, Mohallem NDS (2016) A comparison of TEM and DLS methods to characterize size distribution of ceramic nanoparticles. J Phys Conf Ser 733:012039(5). https//doi.org/10.1088/1742-6596/733/1/012039
32. Khlebtsov BN, Khlebtsov NG (2011) On the measurement of gold nanoparticle sizes by the dynamic light scattering method. Colloid J 73:118–127. https//doi.org/10.1134/S1061933X11010078